# A Frequency Domain Constraint for Synthetic and Real X-ray Image Super Resolution

Qing Ma[1], Jae Chul Koh[2] and WonSook Lee[1]

[1] University of Ottawa, Ottawa, Canada
[1] Korea University Anam Hospital, Seoul, Korea
`{qma088,wslee}@uottawa.ca`

**Abstract.** Synthetic X-ray images are simulated X-ray images projected from CT data. High-quality synthetic X-ray images can facilitate various applications such as surgical image guidance systems and VR training simulations. However, it is difficult to produce high-quality arbitrary view synthetic X-ray images in real-time due to different CT slice thickness, high computational cost, and the complexity of algorithms. Our goal is to generate high-resolution synthetic X-ray images in real-time by upsampling low-resolution images with deep learning-based super-resolution methods. Reference-based Super Resolution (RefSR) has been well studied in recent years and has shown higher performance than traditional Single Image Super-Resolution (SISR). It can produce fine details by utilizing the reference image but still inevitably generates some artifacts and noise. In this paper, we introduce frequency domain loss as a constraint to further improve the quality of the RefSR results with fine details and without obvious artifacts. To the best of our knowledge, this is the first paper utilizing the frequency domain for the loss functions in the field of super-resolution. We achieved good results in evaluating our method on both synthetic and real X-ray image datasets.

**Keywords:** Synthetic X-ray, Super Resolution, Frequency Domain, Digital Reconstructed Radiographs

## 1    Introduction

Efforts have been made on converting CT volume or slice images into synthetic X-ray images, also known as Digital Reconstructed Radiographs (DRRs), with fast rendering algorithms such as ray tracing or ray casting methods [1–4]. In recent years, deep learning aided algorithms can generate highly realistic X-ray images but require vast training data and high computation resources [5, 6]. In general, there are three challenges to high-quality synthetic X-ray image generation. Firstly, it demands high computation resources to produce high-quality synthetic X-ray images, preventing the use in clinics or hospitals. Secondly, the algorithms for generating high-quality synthetic X-ray images are often time-consuming or complex. Finally, the thickness of the scan is not thin enough due to high dose radiation of CT scanning. The gaps between slices yield lower resolution in certain views. A basic requirement for generating fine synthetic X-ray images is that the resolution is approximately equal in all views [7]. Thus, it is necessary to up-sample the lower resolution views, either with interpolation or more sophisticated methods such as semantic interpolation [8] or super-resolution (SR) methods.

Various application scenarios have been explored with synthetic X-ray images. For example, synthetic X-ray images are used in virtual reality (VR) training simulation for training physicians on fluoroscopy-guided intervention procedures [9, 10]. It intends to replace real-life training sessions which are costly and exposed to radiation. High-quality DDRs can also be used to train deep learning models for diagnostic tasks [5]. It's also been found that using synthetic X-ray images can reduce up to half the amount of fluoroscopic images taken during real fluoroscopy-guided intervention procedures [11]. The performance of the above applications could all benefit from higher-quality synthetic X-ray image data sources.

Deep learning-based SR methods have been well explored for natural images. There are two main categories which are Single Image Super-Resolution (SISR) and Reference-based super-resolution (RefSR). SISR aims to reconstruct a high-resolution (HR) image from a single low-resolution (LR) image [12]. RefSR works on learning finer texture details from a given reference HR image [13]. Convolution neural networks (CNN) and Generative adversarial network (GAN) methods have been widely used in SISR. CNN models can reach high performance on evaluation metrics but produce overly smooth images with coarse details. On the other hand, GAN models generate appealing images with fine details but with more artifacts. This issue is especially crucial in medical imaging as image details have an impact on diagnosis and decision-making during operations. RefSR can learn fine texture details from the reference image but still generate some artifacts and noise. Fourier transform is rarely utilized in the field of super-resolution. We propose to use a frequency domain loss as a constraint to mitigate this problem.

We used a matrix-based projection algorithm and custom-built lookup tables created with tissue radiographic opacity parameters to generate fine LR synthetic X-ray images from CT slices in arbitrary views. And then, we use a RefSR method combined with our frequency domain loss to generate HR synthetic X-ray images with few artifacts and noise. We achieved good results on both synthetic and real X-ray image super-resolution with our proposed TTSR-FD.

## 2    Related Work

Deep learning was first used in SISR by Dong et al. [14]. They proposed SRCNN achieved superior results compared to previous conventional SR methods. Lim et al. [12] proposed EDSR introduced Residual Block into SISR that further boosts the model performance. Zhang et al. [15] added the attention mechanism to improve the network performance. These methods used CNN yield a strong PSNR performance. However, the results do not have a good visual quality for human perceptions. GAN gradually become successful in the field known for good visual quality. Ledig et al. [16] proposed SRGAN first adopted GAN and showed appealing image quality. Wang et al. [17] introduced Residual-in-Residual Dense Block and relativistic GAN further improved the perceptual quality of the results.

RefSR takes advantage of learning more accurate texture details from the HR reference image. The reference image could be selected from adjacent frames in a video, images from different viewpoints, etc. [18]. It can achieve visually appealing results without generating many artifacts and noise compared to GAN-based SISR. Zheng et al. [19] proposed CrossNet that adopted a flow-based cross-scale warping to transfer features. SRNTT [18] adopted patch matching with VGG extracted features and can use arbitrary images as references. Yang et al. [13] proposed TTSR applied a transformer network with attention models that outperformed traditional SISR methods. Nevertheless, these models still generate perceivable artifacts and noise in the resulting image.

Fourier transform is a powerful tool in the field of signal processing, and it has also shown great potential in deep learning-related research. Souza and Frayne [20] proposed a hybrid framework for magnetic resonance (MR) image reconstruction that learns in both frequency and spatial domain. Li et al. [21] proposed the first neural network SR method by solely learning in the frequency domain. It shows advantages on the speed of the model but with an imperceptible loss on the quality of results. Xu et al. [22] proposed a novel learning-based frequency channel selection method that achieved superior results on multiple tasks. In this work, we introduce to compute a loss function in frequency domain as a constraint for the neural network, instead of transferring the network itself into frequency domain.

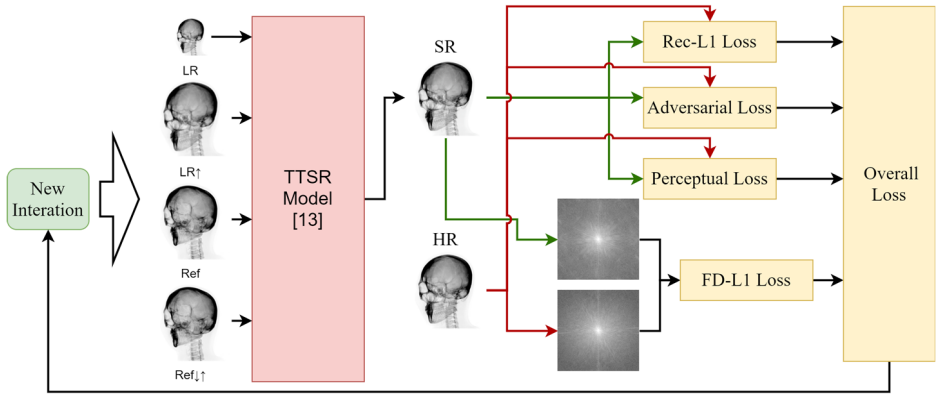

**Fig. 1.** Overview of our TTSR-FD method. We add the frequency domain loss on the TTSR [13] model. The four input images are LR, up-sampled LR, reference images (Ref) and down/up-sampled Ref Image. These rescaling operations will help to improve the texture transfer accuracy [13]. SR image is the model output. HR image is the ground truth image.

## 3 Methods

We aim to reduce the artifacts generated by the network while not losing fine details. To achieve this, we made a thoughtful analysis in the frequency domain for SR methods and introduce frequency domain loss with TTSR [13] as illustrated in Fig. 1. We first explore the frequency domain pattern for images of different quality. Then, we introduce our loss function computed in the frequency domain.

### 3.1 Frequency domain analysis

We adopt Fourier transform to reveal frequency patterns that are not visible in the spatial domain. Different image samples are shown in both spatial and frequency domain in Fig. 2. While the bicubic interpolated image shows the worst results, we observe high-frequency details are also not fully learned by any of the SR models. We show our TTSR-FD result as a comparison of the improvement.

| GT | Input | Bicubic | RCAN | ESRGAN | TTSR | TTSR-FD |

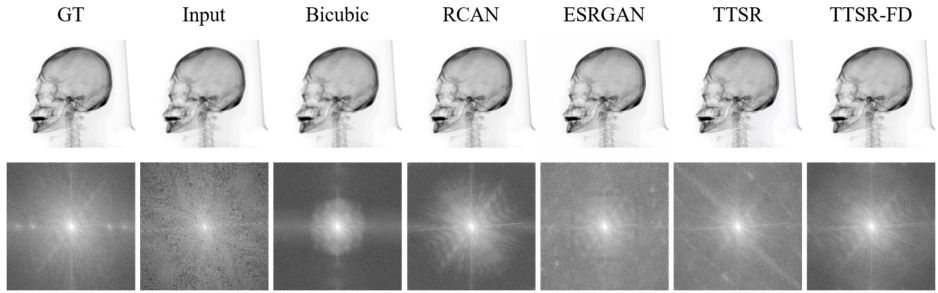

**Fig. 2.** Images in the both spatial and frequency domain from left to right are ground truth, LR input, Bicubic upsampling, and RCAN, ESRGAN, TTSR, TTSR-FD SR methods results.

In general, we can see whiter regions in the center indicating images contain more low-frequency content. The input image loses lots of high-frequency details compare to the GT. The bicubic upsampling is not able to produce any high-frequency details. Limited high-frequency details are generated by RCAN. ESRGAN can learn some good high-frequency details and shows sharp details in the spatial domain. But we can see abnormal patterns in frequency domain such as a rectangle shape contour in the center. This indicates it generates artifacts to satisfy human perceptual. TTSR can generate fine high-frequency details but also fails to match some patterns. Our TTSR-FD got the best similarity in frequency domain compared to GT. We can find out that the frequency domain contains lots of information that is not shown in the spatial domain.

### 3.2 Frequency Domain Loss

We make a hypothesis based on the observation from the previous section that a loss function computed in the frequency domain can increase the resulting image quality of RefSR. We aim to use this loss to set up a constraint during training that can force the model to learn more from the data and generate fewer artifacts and noise. We choose to build it with a pixel-wise loss function to ensure the network learns from the frequency domain patterns. We utilize $L1$ loss which is proven effective for SR methods, also more robust and easier to converge compared to $L2$ loss. Our network architecture is shown in Fig. 1. In short, we compute the loss by transferring model output and ground truth images of each iteration into frequency domain, calculate their $L1$ loss and feed back to the network. We applied this loss to our baseline model TTSR [13], which has three loss functions. The overall loss of our TTSR-FD model is:

$$\mathcal{L}_{overall} = \lambda_{rec}\mathcal{L}_{rec} + \lambda_{fd}\mathcal{L}_{fd} + \lambda_{adv}\mathcal{L}_{adv} + \lambda_{per}\mathcal{L}_{per} \qquad (1)$$

where $\mathcal{L}_{rec}$ is the reconstruction loss with $L1$ loss. $\mathcal{L}_{adv}$ is the adversarial loss using WGAN-GP [23]. $\mathcal{L}_{per}$ is the perceptual loss that include a normal perceptual loss and a texture wise loss [13]. We added a frequency domain loss $\mathcal{L}_{fd}$ to improve the network performance. $\lambda$ is the weight coefficients for the loss functions that are optimized through vast expiriments. In each iteration, a batch of SR and HR images are transferred into the frequency domain and then calculated their $L1$ loss.

$$S^{HR} = f_{rfft}(I^{HR}), \quad S^{SR} = f_{rfft}(I^{SR}) \qquad (2)$$

$$\mathcal{L}_{fd} = \frac{1}{CHW}\|S^{HR} - S^{SR}\|_1 \qquad (3)$$

where $I^{HR}$ and $I^{SR}$ are high resolution and predicted SR result images. $S^{HR}$ and $S^{SR}$ are the corresponding frequency domain images. C, H, W are the channel, height and width of the HR image. $f_{rfft}$ represent the real-to-complex discrete Fourier transform function. We adopt real-to-complex discrete Fourier transform from Pytorch to improve the computation efficiency. We use the built-in FFT function from PyTorch without the need to transfer tensors into other data types. There is no significant increase in time complexity when using our frequency domain loss during training.

## 4 Experiments

### 4.1 Dataset

We create two reference-based SR datasets to evaluate our method, a synthetic head X-ray images dataset and a real chest X-ray images dataset. Both datasets are constructed following similar approaches described in [18] while making some adjustments considering the characteristic of synthetic and real X-ray images.

The synthetic X-ray images dataset is created from head CT data provided by Korea University Anam Hospital. It contains seven different head CT series from five patients with Sagittal (SAG) and Coronal (COR) views. The number of slices in these series ranges from 48 to 145 and each slice resolution is 512x512. We used different linear interpolation variables to fill the gaps between slices for each series. Our projection method is developed in Python with Numba. In the training set, we use a sampling step of 24º for both x and y-axis to generate 225 (15x15) input images for each CT series. The corresponding reference image is generated with a random projection angle range from -45º to 45º. The training images are then cropped into five patches of 160x160 for each input image and reference image. The LR images are generated by downsampling the input images. We build a small and large training set SynXray_S and SynXray_L. Each of them consists of 410 and 1350 image pairs. We choose to leave at least one patient out during training to make sure our test data is not seen during training. The test set images are generated with a sampling step of 20º. There are 30 image groups in the test set from all five patients.

We also create a real chest X-ray images dataset (ChestXrayRef) from ChestX-ray8 [24] dataset for RefSR. We construct the dataset using a similar method in [18]. However, the ChestX-ray8 dataset only has X-ray images in a single frontal view. This makes it impossible to construct the dataset like synthetic X-ray images where the reference image is from a different viewpoint. We choose to randomly pick another chest X-ray image from the dataset as the reference image. This will certainly weaken the ability of RefSR method to reconstruct high-resolution textures, but it is also proven that the reference image doesn't have to be related to the input image [18]. All images are randomly picked and resized to 512 x 512 resolution. The training data has 1298 image pairs. The testing set has 23 image groups.

## 4.2  Training Details

We used a similar training parameter setting as in TTSR [13]. We use Pytorch 1.81 to implement our network. We trained 100 epochs and pick the highest performance model for all datasets. We applied the same parameter setting in all our training. We use a batch size of 4. The learning rate is 1e-4 and Adam optimizer with $\beta_1 = 0.9$, $\beta_2 = 0.999$ and $\epsilon = 1e - 8$. The weight coefficient of the frequency domain loss is 1e-2. The weight coefficient for $\mathcal{L}_{rec}$, $\mathcal{L}_{adv}$, $\mathcal{L}_{per}$ are 1, 1e-3 and 1e-2, respectively. We trained our network for a scaling factor of x4. We use bicubic kernel for rescaling the input and reference images. We trained our model with a Tesla P100 GPU. The training time for large and small synthetic datasets is around 10 and 4 hours respectively. The training time for the real X-ray dataset is around 18 hours.

## 4.3  Results

We first compare our method on synthetic X-ray images dataset with state-of-the-art methods, such as RCAN [15], ESRGAN[17] and TTSR[13]. We make some modifications to our baseline model TTSR for training grayscale images, where we duplicate the grayscale channel into three channels corresponds to RGB channels for natural images then feed to the neural network. We retrain RCAN and ESRGAN models on the paired SISR dataset version of our dataset for comparison purposes as well. We evaluate our method on PSNR and SSIM as shown in Table 1. TTSR-FD achieves superior performance on both metrics. We also observe that TTSR-FD is strong on learning from a small dataset which indicates it can learn more from the data. This can be beneficial for real-life applications since medical data are hard to obtain.

We show the visual comparison for our method trained with SynXray_S on the test set in Fig. 3, consider that the visual difference between large and small datasets is minimal for our method. RCAN shows the second-highest accuracy in the quantitative measurement but we observe the image is overly smoothed and has limited high-frequency details. We can observe that the noise from the red circle area and black dots artifacts from the red rectangle area in TTSR have been removed in TTSR-FD. Our method can generate fine details without generating much noise and artifacts.

**Table 1.** PSNR/SSIM comparison among different SR methods on SynXray_S and SynXray_L datasets. The best results are in bold.

| Method | SynXray_S | SynXray_L |
|--------|-----------|-----------|
| RCAN | 38.597/0.9482 | 38.880/0.9496 |
| ESRGAN | 34.483/0.9023 | 35.270/0.9140 |
| TTSR | 37.953/0.9393 | 38.115/0.9431 |
| TTSR-FD (ours) | **39.009/0.9521** | **39.261/ 0.9514** |

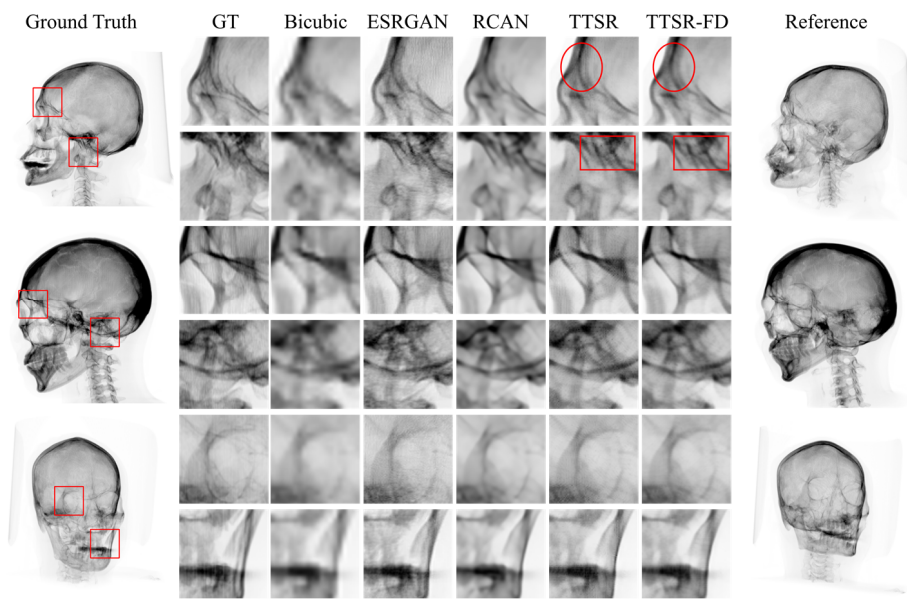

**Fig. 3.** Visual comparison for our method on the testing set. Example input and Ref images are generated from CT series in sagittal (1, 2) and coronal (3) view. TTSR-FD is ours.

We further test our model on Real chest X-ray images. The two models have similar quantitative results as shown in Table 2. However, we find out that the artifacts generated in TTSR-FD are significantly less than in TTSR results as shown in Fig. 4. This validates the effectiveness and generalizability of our proposed method.

**Table 2.** PSNR/SSIM comparison among TTSR and TTSR-FD on ChestXrayRef dataset

| Method | PSNR | SSIM |
|--------|------|------|
| TTSR | **36.767** | 0.9457 |
| TTSR-FD | 36.766 | **0.9470** |

#### 4.4 Ablation study

In this section, we further verify the effectiveness of our proposed method. We set up TTSR-Rec101 that has the same ratio of weight coefficients distribution as TTSR-FD where we only increase the weight coefficients of reconstruction loss $\mathcal{L}_{rec}$ from 1 to 1.01, compared to the weight coefficients of TTSR-FD for $\mathcal{L}_{rec}$ and $\mathcal{L}_{fd}$ are 1 and 0.01. The weight coefficients of other loss functions are the same for both models. Our results are shown in Table 3. The proposed frequency domain loss significantly improved the model performance. We show the resulting images in frequency domain in Fig. 5. We can see that without frequency domain loss, increasing the weight coefficient of the reconstruction loss could not improve the similarity of high-frequency patterns in frequency domain. We observe that the vertical line patterns are not learned by TTSR-rec101 or TTSR as shown in red arrows.

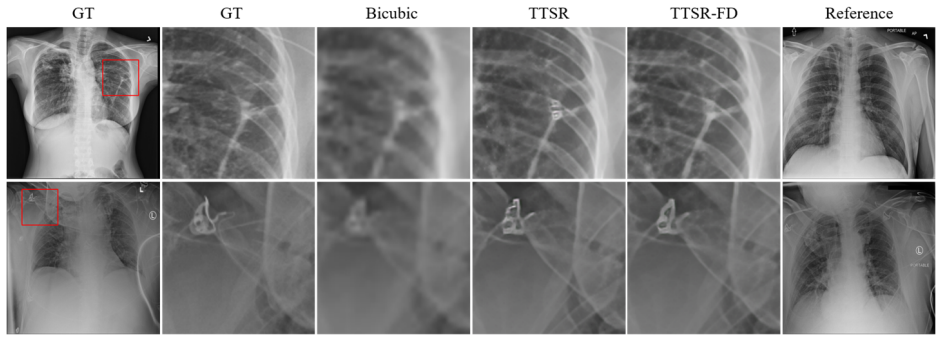

**Fig. 4.** Real Chest X-ray visual comparison between TTSR and TTSR-FD

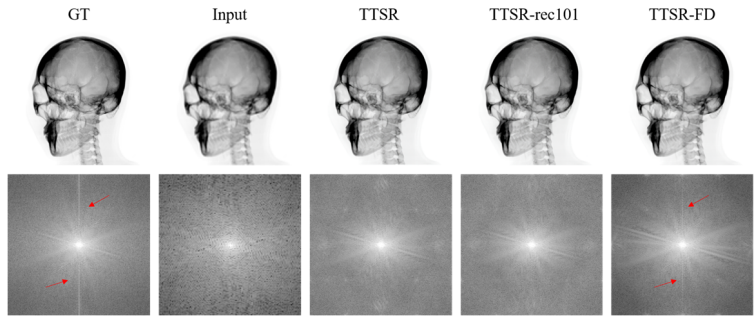

**Fig. 5.** Ablation study on frequency domain loss. Vertical line patterns indicated in red arrows in the ground truth image are only learned by TTSR-FD.

**Table 3.** Ablation study for frequency domain loss with SynXray_S dataset

| Method | PSNR | SSIM |
|---|---|---|
| TTSR | 37.953 | 0.9393 |
| TTSR-Rec101 | 38.245 | 0.9405 |
| TTSR-FD | **39.009** | **0.9521** |

# 5      Conclusion

We proposed a texture transformer super-resolution with frequency domain loss as a constraint for synthetic and real X-ray image super-resolution. We demonstrate that an additional loss function computed in the frequency domain can improve the image quality for synthetic and real X-ray images super-resolution. Our work enables the possibility of generating high-quality synthetic X-ray images in real-time for image guiding systems and VR simulations. Even though the experiments were done on X-ray images, the proposed method is applicable for other medical images. In future works, we would like to explore more possible loss variants and other application scenarios such as denoising with frequency domain loss.